# Decentralized Dynamic State Estimation with Bimodal Gaussian Mixture Measurement Noise

Vahid Sarfi, Amir Ghasemkhani, Iman Niazazari, Hanif Livani, and Lei Yang

*Abstract*—**This paper proposes a decentralized dynamic state estimation (DSE) algorithm with bimodal Gaussian mixture measurement noise. The decentralized DSE is formulated using the Ensemble Kalman Filter (EnKF) and then compared with the unscented Kalman filter (UKF). The performance of the proposed framework is verified using the WSCC 9-bus system simulated in the Real Time Digital Simulator (RTDS). The phasor measurement unit (PMU) measurements are streamed in real-time from the RTDS runtime environment to MATLAB for real-time visualization and estimation. To consider the data corruption scenario in the streaming process, a bi-modal distribution containing two normal distributions with different weights and variances are added to the measurements as the noise component. The performances of both UKF and EnKF are then compared for by calculating the mean-squared-errors (MSEs) between the actual and estimated states.**

*Index Terms*-- **Ensemble Kalman Filter; Dynamic State Estimation; Real-Time Digital Simulator; Unscented Kalman Filter; Phasor Measurement Unit**

## I. INTRODUCTION

State Estimation (SE) is a widely used tool in the power system to monitor the power system states [1]. State estimation is generally divided into two approaches; Static State Estimation (SSE) and Dynamic State Estimation (DSE). SSE approaches, such as weighted least squares (WLS) [2], are utilized in utilities for estimation of power system states, i.e., voltage magnitudes and angles, under steady-state conditions. However, DSE is formulated to estimate generators states, e.g., rotor angle, rotor angle deviation, under dynamic contingencies.

With the proliferation of phasor measurement units (PMUs), with the streaming rate of 30 to 60 samples-per-second (SPS), PMU data-driven applications have become feasible [3],[4] and can be utilized in the utilities to estimate the generator states under dynamic situations. In [5] several DSE algorithms are reviewed and implemented based on PMU data. In [6]-[7], a sample-based Kalman filter known as unscented Kalman filter (UKF) is used to obtain the mean and covariance of the dynamic states of generators. However, the

UKF-based DSE is derived based on the unimodal Gaussian distribution assumption for both state transition and measurement noise that can lead to weak performance in the presence of non-Gaussian measurement noise. Reference [8] shows that measurement noise follows non-Gaussian distribution functions, such as bi-modal distribution containing two normal distributions with different weights and variances. The Ensemble Kalman Filter (EnKF) which is based on Monte Carlo sampling approach, is beneficial for the state-estimation with unknown or non-Gaussian state or measurement noises. In EnKF-based algorithms, Gaussian distributions assumption is utilized for the prior and posterior distributions. However, the Monte Carlo sampling approach helps the algorithm to be more robust compared to the other estimation algorithms such as UKF. Hence, it is of paramount importance to investigate the effects of measurements noise with non-Gaussian distribution on the performance of the DSE algorithms.

Therefore, this paper proposes a decentralized DSE algorithm, based on two different Kalman Filter approaches, UKF and EnKF, to develop and evaluate the impact of bi-modal Gaussian mixture noise on PMU data. This algorithm utilizes the voltage magnitudes and angles, and the power injections that are streamed from a simulated real-time environment, RTDS.

The rest of this paper is organized as follows: In Section II, the decentralized dynamic state estimation framework is presented. Real-time simulation of the DSE framework is discussed in Section III. Section IV introduces the case study and show all the simulation results. Section V summarizes the paper with some concluding remarks.

## II. DECENTRALIZED DYNAMIC STATE ESTIMATION

In this paper, decentralized DSE framework is used for prediction of the generators' states, and then to correct the predicted states using the latest available PMU data. Predicting the states in advance helps in filling the gap in the measurements, arising due to varying communication delays in the networks, and generates the pseudo measurements, which can be used in the measurement update step, in case of incomplete information or the communication link failure.

### A. Power System Model

It is assumed that the generalized non-linear model of the power system can be represented in the following global structure:

$$x(k + 1) = g[x(k), u(k), w(k), k]$$
$$y(k) = h[x(k), u(k), k] + v(k) \qquad (1)$$

This work is partially supported by NSF award, # 1723814.

V. Sarfi, I. Niazazari and H. Livani are with the Department of Electrical and Biomedical Engineering; University of Nevada, Reno (emails: vsarfi@nevada.unr.edu, niazazari@nevada.unr.edu, hlivani@unr.edu). A. Ghasemkhani and L. Yang are with the Department of Computer Science and Engineering; University of Nevada, Reno (emails: aghasemkhani@nevada.unr.edu, leiy@unr.edu).

where $x(k)$ is the state variable vector at time $k$, $u(k)$ is the input vector, $w(k)$ is the process noise, $v(k)$ is the measurement noise which is a bi-modal distribution containing two normal distributions with different weights and variances, $y(k)$ is the measurable output, and $g$ and $h$ are the system and output functions respectively.

Since the decentralized DSE is the focus of this paper, the general power system model in (1) can be simplified to its equivalent form of a single machine connected to an infinite bus via transmission lines where the generator states will be considered as the states of the system [9]. This equivalent dynamic model will be the basis for developing and validating our generator state estimator in RTDS. The generator dynamic state-space model can then be characterized by the following fourth-order nonlinear equations [10]-[11]:

$$x = [\delta\ \Delta\omega\ e_q'\ e_d']^T = [x_1\ x_2\ x_3\ x_4]^T$$

$$u = [T_m\ E_{fd}\ ]^T = [u_1\ u_2]^T$$

$$y = [P_t\ Q_t\ ]^T = [y_1\ y_2]^T \qquad (2)$$

$$\dot{x}_1 = \omega_0 x_2$$

$$\dot{x}_2 = \frac{1}{J}(u_1 - T_e - Dx_2)$$

$$\dot{x}_3 = \frac{1}{T_{do}'}(u_2 - x_3 - (x_d - x_d')i_d)$$

$$\dot{x}_4 = \frac{1}{T_{qo}'}(-x_4 + (x_q - x_q')i_q)$$

$$y_1 = P_t = T_e = \frac{V_t}{x_d'}x_3 \sin x_1 + \frac{V_t^2}{2}\left(\frac{1}{x_q} - \frac{1}{x_d'}\right)\sin 2\,x_1$$

$$y_2 = Q_t = \frac{V_t}{x_d'}x_3 \cos x_1 - V_t^2\left(\frac{\sin^2 x_1}{x_q} + \frac{\cos^2 x_1}{x_d'}\right) \qquad (3)$$

where $\delta$ is the rotor angle in (elec.rad), $\omega$ is the rotor speed in (pu), $e_q'$ and $e_d'$ are $q-axis$ and $d-axis$ transient voltages in (pu), respectively, $\omega_0 = 2\pi f_0$ is the nominal synchronous speed in (elec.rad/s), $T_e$ is the air-gap torque which is assumed to be equal to electrical active power (i.e. $P_t$) in (pu), and $T_m$ and $E_{fd}$ are the mechanical input torque in (pu) and the exciter output voltage in (pu), respectively. Besides, $P_t$ and $Q_t$ are the system's active and reactive output powers which are assumed to be the system's measurements. The other variables are the nominal data of the selected synchronize machine.

The DSE scheme is shown in Fig.1. As it can be seen, for predicting the states in the next time step using the nonlinear models, we need the PMU data (active and reactive power injection) at the generator bus, generator's parameters, and $E_{fd}$ and $T_m$ input to the generator [10].

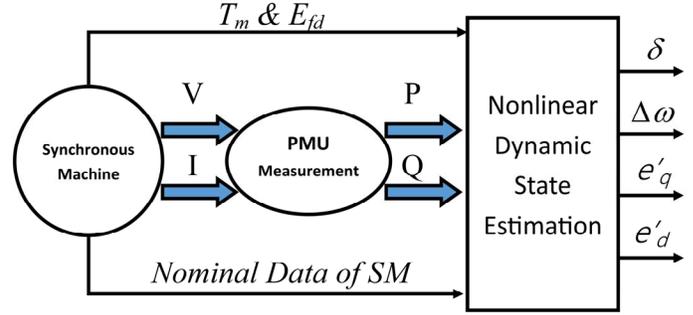

Fig.1. Decentralized DSE scheme

### B. Unscented Kalman Filter (UKF)

The UKF is a sample-based estimation method, which leverages the unscented transformation using a deterministic sampling approach to approximate the mean and covariance of a random state vector. The minimum number of samples are selected and then propagated into the system's non-linear functions to calculate the new mean and covariance of the output states. However, UKF algorithm assumes a unimodal Gaussian distribution for both state transition and measurements noises, which can be imprecise in the case of multimodality in the system's noises distributions [12]. The results in this paper show that with non-Gaussian measurement noise, UKF performs poorly to predict and update the system states.

### C. Ensemble Kalman Filter (EnKF)

EnKF is a sampling-based estimation approach, which was first introduced by Evensen [13] as a data assimilation method. EnKF uses Monte Carlo sampling approach to generate the particles and assumes Gaussian distributions for the prior and posterior distributions. Zhou et al. demonstrate that the EnKF algorithm outperforms other algorithms when the typical PMU sampling rate is used for estimation and is more compatible with online implementation [14]. The provided examples in [14] show that the EnKF yields better results for dynamic state estimation in the power system, which is a complex and highly non-linear problem. However, to the best of the authors' knowledge, the performance of the EnKF against unknown or non-Gaussian noises has not been properly studied. In this paper, we would like to assess the robustness of EnKF-based DSE with respect to non-Gaussian measurement noise and compare it with UKF-based method.

### D. Measurement Characteristics

Despite the advantages of PMUs, the captured data stream might have some quality issues such as large measurement bias, non-Gaussian noise, data packet loss, loss of GPS synchronization, measurement delays, missing timestamps, and cyber-attack. The non-Gaussian PMU noises could follow Laplace, Cauchy, bimodal or tri-modal Gaussian mixture distributions. Fig.2 shows two different bimodal Gaussian mixture distributions with either zero means or various means.

There are several models that can be applied to model deviations from the Gaussian assumption. Among them, the Gaussian sum distribution is widely used because any non-Gaussian distribution *p(x)* can be expressed as, or approximated sufficiently well, by a finite sum of known Gaussian densities according to the Wiener approximation theorem [15]

$$p(x) = \sum_{i=1}^{N_c} w_i N(\mu_i, \sigma_i^2) \qquad (4)$$

where $w_i$ is the weight and $\sum_{i=1}^{N_c} w_i = 1$; $N_c$ is the number of Gaussian components; the mean and covariance matrix associated with the $i^{th}$ Gaussian component are denoted by $\mu_i$ and $\sigma_i^2$, respectively.

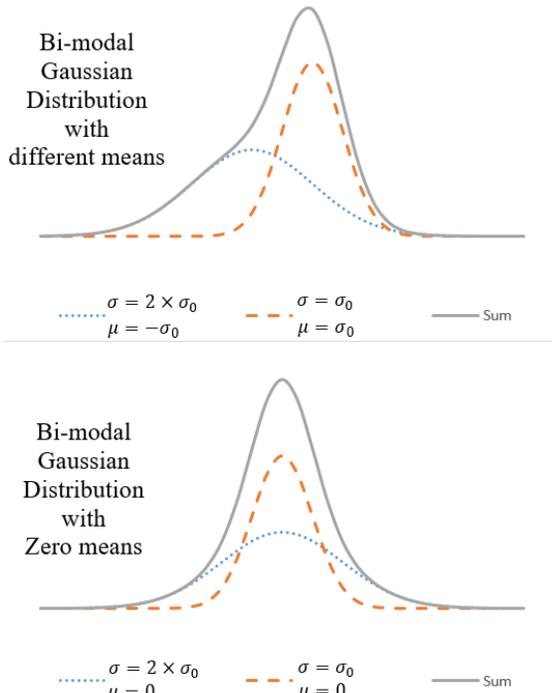

Fig.2. Bimodal Gaussian mixture distributions with either zero means or various means

In this paper, we study the effect of bimodal Gaussian mixture measurement noise as one of the common non-Gaussian noise types, on the performance of the decentralized DSE.

## III. REAL-TIME SIMULATION OF DSE

### A. RTDS simulation

RTDS is a real-time simulation platform, which uses a combination of hardware and software components to carry out power system simulations in real-time [16]. The Gigabit Transceiver Network (GTNET) card is utilized to connect the RTDS to External devices. The RTDS, used for the present work, is equipped with the GTSYNC card to receive the one pulse per second (pps) signal from the satellite through GPS clock. This signal is used to synchronize the software PMUs with the GPS clock.

The GTNET card supports various protocols including the IEEE C37.118-2011 protocol [17] to send PMU data to executing the DSE as shown in Fig. 3. To acquire the PMU data, the output of the GTNET is received by the MATLAB through the communication channel. To capture the measurements at various operating conditions, load variation commands are sent to the RTDS from the DSE application platform at regular intervals.

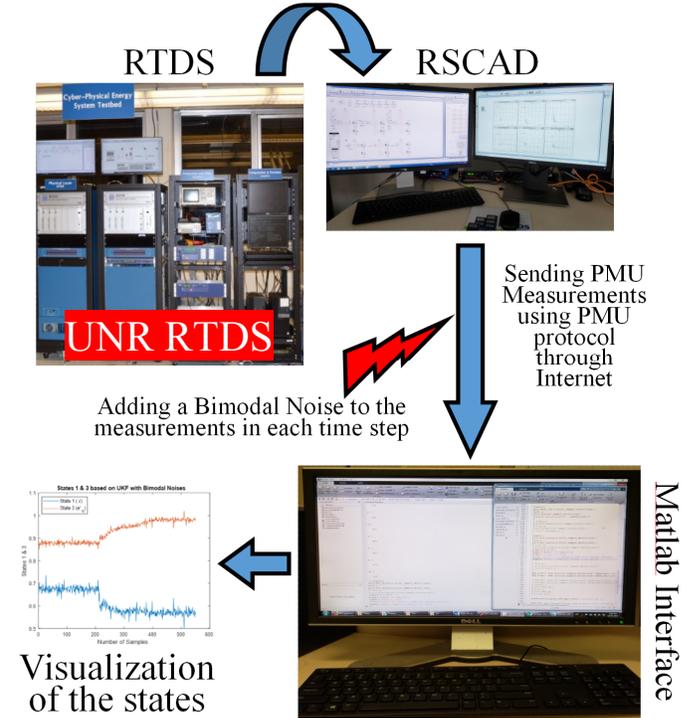

Fig.3. The architecture of the interface between RTDS and MATLAB

### B. DSE simulation

In this paper, the GTNET PMUs are configured to provide measurements, active and reactive power injections, at the rate of 30 or 60 samples-per-second (SPS). The UKF- and EnKF-based DSE algorithms are executed, normally at the rate of PMU measurement data refresh, by utilizing the latest available data. Using the Software in the Loop (SIL) configuration, the DSE execution process is repeated for numerous loading conditions by automatically changing the load of the test system. In this paper, the UKF- and EnKF-based DSE algorithms are carried out with Gaussian mixture measurement noises.

## IV. EXPERIMENTAL RESULTS AND DISCUSSION

### A. Case Study

Figure 4 shows the WSCC 9-bus system diagram that is simulated in RTDS and then augmented with an automated script in MATLAB to capture the PMU data and execute the DSE algorithms. Noted that for all the simulations, the experimental platform is Win7; Intel with CPU E5420 2×2.5 GHz; RAM 16 GB.

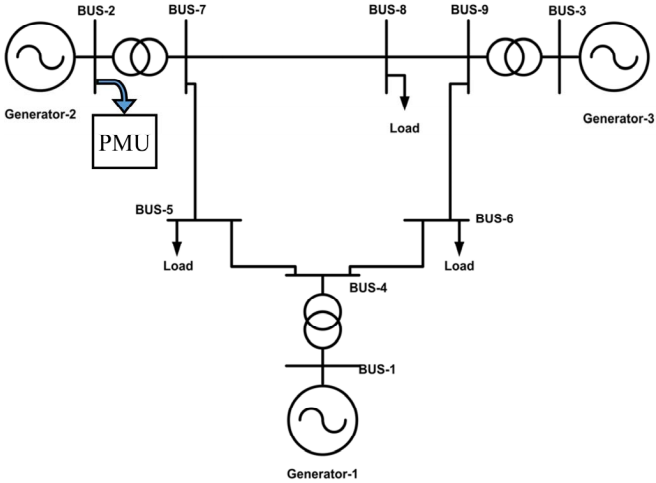

Fig.4. The WSCC 9-bus system diagram

### B. Simulation Results and discussion

The PMU data obtained by the RTDS are noise-free, compared to actual PMU data from the field. To better match the obtained RTDS data with the world PMU measurements, a bi-modal Gaussian mixture noise with zero means, variances of $10^{-4}$ and $10^{-3}$, and weights of 0.9 and 0.1, is added to active and reactive power measurements [8]. We then use the noisy data to evaluate the performance of the UKF- and EnKF-based DSE approaches under the Gaussian mixture noise scenarios. The estimated states (the rotor angle, the rotor speed, and q-axis and d-axis transient voltages) are compared with the actual values to determine the accuracy of the above approaches with non-Gaussian measurement noise. In this paper, the PMU streaming rate is 60 SPS and the second generator's dynamic states are estimated and compared for accuracies calculation. In order to assess the performance of the DSE algorithms under a dynamic contingency scenario, the load at bus 5 is disconnected at t=3.5 *sec*. It must be noted that execution time for EnKF-based method with 100 sampling points and UKF-based method is around 3 *msec* which is less than the time interval between two subsequent PMU data, 16.6 *msec*. Therefore, both methods are able to utilize the latest available PMU data to estimate the dynamic states in an online manner.

Fig. 5 and Fig. 6 show the estimated dynamic states 1 and 3 (rotor angle and q-axis transient voltage) for UKF-based and EnKF-based DSE methods with the considered measurement noise. From the existing literature, we know that UKF- and EnKF-based DSE methods perform well with Gaussian measurement noise and it is difficult to distinguish the superiority of one method over the other one. As can be seen in Fig. 5, UKF-based method performs poorly with non-Gaussian PMU measurement noise. However, as can be seen in Fig. 6, EnKF-based method shows superior performance compared to the UKF-based method with non-Gaussian noise, such as bimodal Gaussian mixture model.

In addition, Fig. 7 and Fig. 8 show the square error of state 3 for both UKF- and EnKF-based method, respectively. As it can be seen, for each sampling instant, the EnKF-based method leads to smaller errors compared to the UKF-based method, and subsequently a smaller Mean Square Error (MSE). It must be noted that errors and MSE trends are the same for all other states. The calculated MSEs for all four states for both methods are shown in Table I.

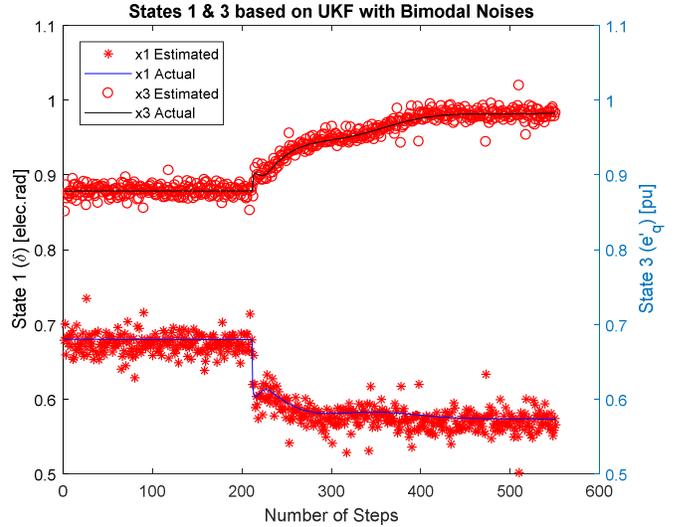

Fig.5. State estimation of $x_1 = \delta$ and $x_3 = e'_q$ using UKF based DSE process with PMU data with bi-modal Gaussian mixture noise

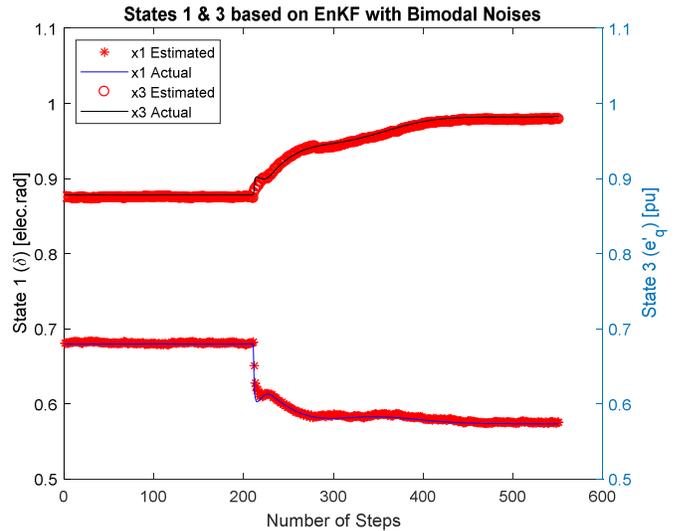

Fig.6. State estimation of $x_1 = \delta$ and $x_3 = e'_q$ using EnKF based DSE process with PMU data with bi-modal Gaussian mixture noise

The reason behind the weak performance of the UKF is that this method is designed to work with Gaussian assumptions for measurement or state transition noises by creating several sigma points around the mean using the unscented transformation. However, with non-Gaussian noises, it has difficulty calculating the states in both prediction and update steps. On the other hand, EnKF uses the Monte Carlo simulation to generate the points in each iteration and it does work with any specific probability distribution. The important parameter that affects the results of EnKF is the number of Monte Carlo sampling. The estimation accuracy increases with an increase in the number of samples, with an increase in the computation time as well. The trade-off between the accuracy and computation time is defined by the user. In this paper, we use 100 sampling points to achieve the desired accuracy while keeping the execution time below 16 *msec* threshold.

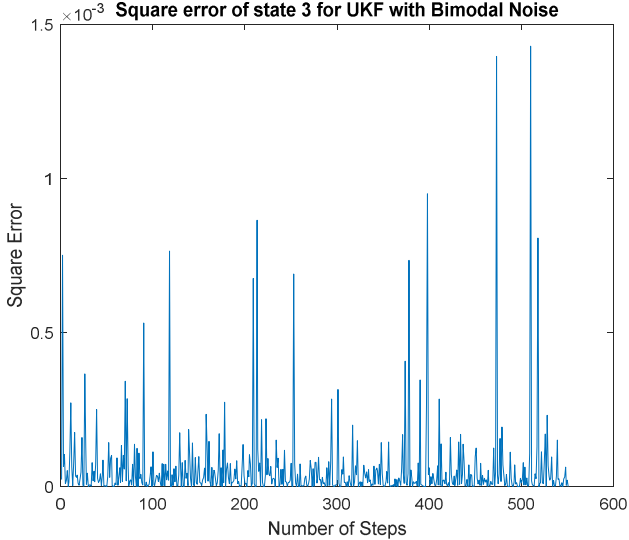

Fig.7. Square error of state 3 $x_3 = e'_q$ with respect to its real values using UKF based DSE process with PMU data corrupted by bi-modal Gaussian mixture noise

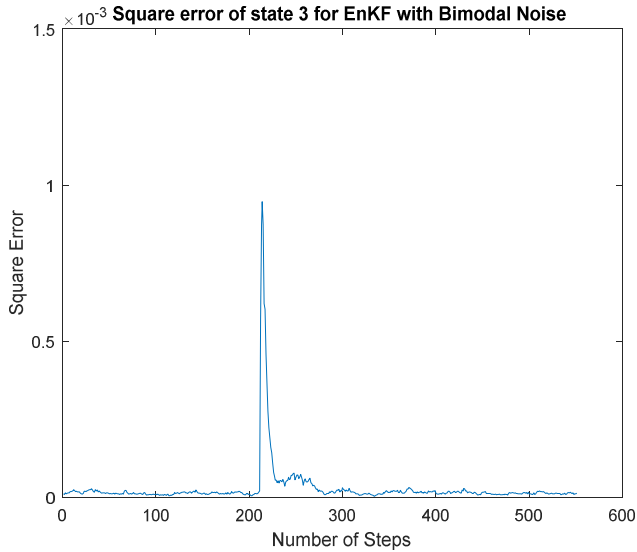

Fig.8. Square error of state 3 $x_3 = e'_q$ with respect to its real values using EnKF based DSE process with PMU data corrupted by bi-modal Gaussian mixture noise

Table I. MSE for all four dynamic states

|  | x1 | x2 | x3 | x4 |
|---|---|---|---|---|
| **UKF** | 5.11×10⁻⁵ | 4.95×10⁻⁵ | 5.76×10⁻⁵ | 5.2×10⁻⁵ |
| **EnKF** | 1.02×10⁻⁵ | 1.68×10⁻⁵ | 1.87×10⁻⁵ | 1.13×10⁻⁵ |

## V. Conclusion

This paper compares the performance of the Unscented Kalman filter (UKF) and the Ensemble Kalman Filter (EnKF) algorithm for decentralized dynamic state estimation (DSE) with bimodal Gaussian mixture measurement noise. The performance of the proposed framework is tested by estimating rotor angle, rotor speed, $q-axis$ and $d-axis$ transient voltages of the generators in the WSCC 9-bus system. This estimation is solved subject to load disconnection dynamic contingencies in the presence of a bi-modal Gaussian mixture noise. The estimation results verify the superior performance of the EnKF-based method compared to the UKF-based method. These results include mean square error for various states of the generators with different noise characteristics.